\documentclass[12pt,a4paper]{article}
\usepackage{natbib}
\usepackage{multirow} 
\usepackage{url}
\usepackage{array}
\usepackage{bm}
\usepackage{epsfig}
\usepackage{graphicx}

\usepackage[latin1]{inputenc}
\usepackage{amsmath,graphicx}

\usepackage{comment}

\title{Multiple imputation and selection of ordinal level 2 predictors in multilevel models. An analysis of the relationship between student ratings and teacher beliefs and practices.}

\author{Leonardo Grilli (University of Florence)
\and
 Maria Francesca Marino (University of Florence)
\and
Omar Paccagnella   (University of Padua)
\and
 Carla Rampichini  (University of Florence)}

\date{ }
    
\begin{document}
\maketitle

\begin{abstract}
The paper is motivated by the analysis of the relationship between student ratings and teacher practices and beliefs, which are measured via a set of binary and ordinal items collected by a specific survey with nearly half missing respondents. The analysis, which is based on a two-level random effect model, must face two issues about the items measuring teacher practices and beliefs:  (\textit{i}) these items  are level 2 predictors severely affected by missingness; (\textit{ii}) there is redundancy in both the number of items and the number of categories of their measurement scale. We tackle the first issue by considering a multiple imputation strategy based on information at both level 1 and level 2. For the second issue, we consider regularization techniques for ordinal predictors, also accounting for the multilevel data structure. The proposed solution combines existing methods in an original way to solve the specific problem at hand, but it is generally applicable to settings requiring to select predictors affected by missing values. The results obtained with the final model point out that some teacher practices and beliefs are significantly related to ratings about teacher ability to motivate students. 

\vspace{1em}
\noindent Keywords: Lasso, MICE, Random effects, University course evaluation.

\end{abstract}
\vspace{-8pt}

\section{Introduction}
\label{sec:intro}

The evaluation of university courses, which is essential for quality insurance, is typically based on student ratings. A large body of literature focuses on studying the factors associated with the expressed evaluations, such as the characteristics of the student, the teacher and the course \citep{spooren2013}. It is widely recognized that teaching quality is a key determinant of student satisfaction, even if teacher observed characteristics often reveal weak effects \citep{hanushek2006}. Therefore, it is helpful to gather more information about teacher practices and beliefs by specific surveys involving the teachers themselves \citep{goe2008}. In this vein, the PRODID project, launched in 2013 by the University of Padua \citep{Dalla Zuanna:16}, is a valuable source as it implemented a new CAWI survey addressed to the teachers for collecting information on their practices and beliefs about teaching. 

We aim at analysing the relationship between student ratings and teacher practices and beliefs, controlling for the available characteristics of students, teachers and courses. Given the hierarchical structure of ratings nested into teachers, we exploit multilevel modelling \citep{Goldstein:2010, rampichini2004}. Teacher practices and beliefs from the PRODID survey enter the model as level 2 predictors, but they are missing for nearly half of the teachers due to non-response. Thus, the multilevel analysis must face a serious issue of missing data at level 2. This issue is receiving increasing attention in the literature \citep{Grund:18}. In addition, modelling the effects of teacher practices and beliefs is complicated since they are measured by a wide set of binary and ordinal items, calling for suitable model selection techniques. Therefore, the case study raises the methodological challenge of selecting level 2 predictors affected by missing values.
We handle the missing values through multiple imputation by chained equations, exploiting information at both level 1 and level 2 \citep{Grund:17, Mistler:17}. For the selection of predictors, we consider regularization techniques for ordinal predictors \citep{Gertheiss:10} and we propose a strategy to combine selection of predictors and imputation of their missing values.

The rest of the paper is organized as follows. Section \ref{sec:data} describes the data structure and the model specification. Section \ref{sec:imputation} outlines the imputation procedure to handle missing data at level 2, then Section \ref{sec:lasso} presents the regularization method chosen to deal with ordinal predictors.  Section \ref{sec:combine} outlines the proposed strategy to combine imputation and model selection, while Section \ref{sec:results1} illustrates the application of the strategy to the case study, reporting the main results. Section \ref{sec:conclusion} concludes with some remarks and directions for future work.

\section{Data description and model specification}
\label{sec:data}

As anticipated in Section \ref{sec:intro}, we wish to analyse the relationship between student ratings and teacher practices and beliefs, controlling for the available characteristics of the student, the teacher, and the course. To this end, we exploit a dataset of the University of Padua for academic year 2012/13, obtained by merging three sources: (\emph{i}) the traditional course evaluation survey with $18$ items on a  scale from 1 to 10, where 10 is the maximum; (\emph{ii}) administrative data on students, teachers, and courses; (\emph{iii}) the innovative PRODID survey collecting information on teacher practices and beliefs \citep{Dalla Zuanna:16}. 

Data have a two-level hierarchical structure, with $56,775$ student ratings at level 1 and $1,016$ teachers at level 2. The average group size is $79$ (min 5, max 442). 

We investigate student opinion about teacher ability to motivate students, which is one of the items of the course evaluation questionnaire\footnote{https:$\setminus\setminus$www.unipd.it$\setminus$opinione-studenti-sulle-attivita-didattiche}.
The analysis is based on the following 2-level linear model for rating $i$ of teacher $j$:
\begin{equation}
\label{eq:model}
y_{ij}=\mathbf{x}_{ij}'\bm\alpha +  \mathbf{z}_{j}' \bm\delta +  \mathbf{q}_{j}'\bm\gamma +u_{j}+e_{ij},
\end{equation}
where $\mathbf{x}_{ij}$ is the vector of level 1 covariates (student characteristics) including the constant, while $\mathbf{z}_{j}$ is the vector of fully observed level 2 covariates (administrative data on teachers and courses), and $\mathbf{q}_{j}$ is the vector of partially observed level 2 covariates (teacher practices and beliefs). 
Model errors are assumed independent across levels with the standard distributional assumptions, namely $e_{ij} \stackrel{\hbox{\tiny iid}}{\hbox{$\sim$}} N(0, \sigma^2_e)$ and  $u_{j} \stackrel{\hbox{\tiny iid}}{\hbox{$\sim$}} N(0, \sigma^2_u)$. 

The survey on teacher beliefs and practices has about fifty percent of missing questionnaires, posing a serious issue of missing data at level 2. An analysis based on listwise deletion would discard the entire set of student ratings for non responding teachers, causing two main problems: (\emph{i}) a dramatic reduction of the sample size, and thus of the statistical power, and (\emph{ii}) possibly biased estimates if the missing data mechanism is not MCAR. To overcome these issues, we impute missing values by means of multiple imputation, which allows us to retain all the observations and to perform the analysis under the more plausible MAR assumption \citep{Seaman:13}.

\section{Handling missing data at level 2}
\label{sec:imputation}
In multilevel models, the treatment of missing data requires special techniques since missing values can occur at any level of the hierarchy. Furthermore, missing values can alter variance components and correlations. 

Multiple imputation (MI) is a flexible approach to handle missing data taking into account the uncertainty deriving from the  imputation procedure. MI is carried out in two steps: (\textit{i}) generate several imputed data sets according to a suitable imputation model; (\textit{ii}) fit the substantive model on each imputed data set, and join the results using Rubin rules \citep{Little:02}. 
The two main approaches to implement MI are \textit{Joint Modelling} (JM) and fully conditional specification, also known as \textit{Multivariate Imputation by Chained Equations} (MICE), see \cite{vanBuuren:18} for a comprehensive treatment, and  \cite{Mistler:17, Grund:17} for a comparison of these approaches in multilevel settings. In the JM approach, data are assumed to follow a joint multivariate distribution and imputations are generated as draws from the fitted distribution. In the MICE approach, missing data are replaced by iteratively drawing from the fitted conditional distributions of partially observed variables, given the observed and imputed values of the remaining variables in the imputation model. 
In our case missing data are only at level 2, so we can apply MI techniques to the level 2 data set and then merge level 1 and level 2 data sets. According to the literature on MI in multilevel settings,  the imputation model used to simulate missing information at level 2 should include level 2 covariates, the cluster size, and proper summaries of level 1 variables (covariates and response variable). Several strategies may be adopted to summarize level 1 variables: if all the variables are normal, the sample cluster mean is the optimal choice \citep{Carpenter:13}. For the imputation of categorical variables there are no theoretical results, but simulation studies show that the sample cluster mean is a good compromise between accuracy and computational speed \citep{Erler:16, Grund:18}. Therefore, we summarise level 1 variables through the cluster mean, which is easy to implement in our case since level 1 variables are completely observed. 

In our case the imputation step is challenging since we have to impute many categorical variables, indeed about 50\% of the teachers did not respond to the whole questionnaire producing missing values on $10$ binary items (teacher practices) and $20$ ordinal items (teacher beliefs on a 7-point scale). The JM approach is computationally demanding with many ordinal items, thus we rely on the MICE approach, performing imputations using the \texttt{mi chained} command of Stata \citep{Stata:17}. 
The imputation model is composed of binary logit models for the $10$ binary items (teacher practices) and cumulative logit models for the $20$ ordinal items (teacher beliefs). The imputation model includes the following fully observed covariates:  teacher characteristics, course characteristics (including the number of ratings), and the cluster means of the ratings for all questions of the course evaluation questionnaire, including the response variable. The inclusion of mean ratings increases the plausibility of the MAR assumption.

\section{Selecting ordinal predictors with regularization techniques}
\label{sec:lasso}
The PRODID questionnaire measures teacher practices using  10 binary items and teacher beliefs using 20 ordinal items on a 7-point Likert scale. Such items contain information on a few dimensions of teaching that  in principle could be summarized using latent variable models for ordinal items \citep{Bartholomew:11}. However, about $50\%$ of the teachers did not respond to the questionnaire, thus applying latent variable methods to the complete cases can lead to biased results. On the other hand, fitting latent variable models using imputed data sets raises two main problems: (\emph{i}) how to combine the results in order to identify the latent dimensions and assign the corresponding  scores to the teachers, and (\emph{ii}) how to take into account the variability of the predicted scores in the main model. 
The literature on factor analysis in the presence of missing responses is growing \citep{Seva:16, Nassiri:18}, but the issue is still controversial, thus we prefer to directly use the imputed PRODID items as covariates in the main model and select them applying model selection techniques.

\begin{figure}[h!]
\centering
\makebox{\includegraphics[width=0.8\textwidth]{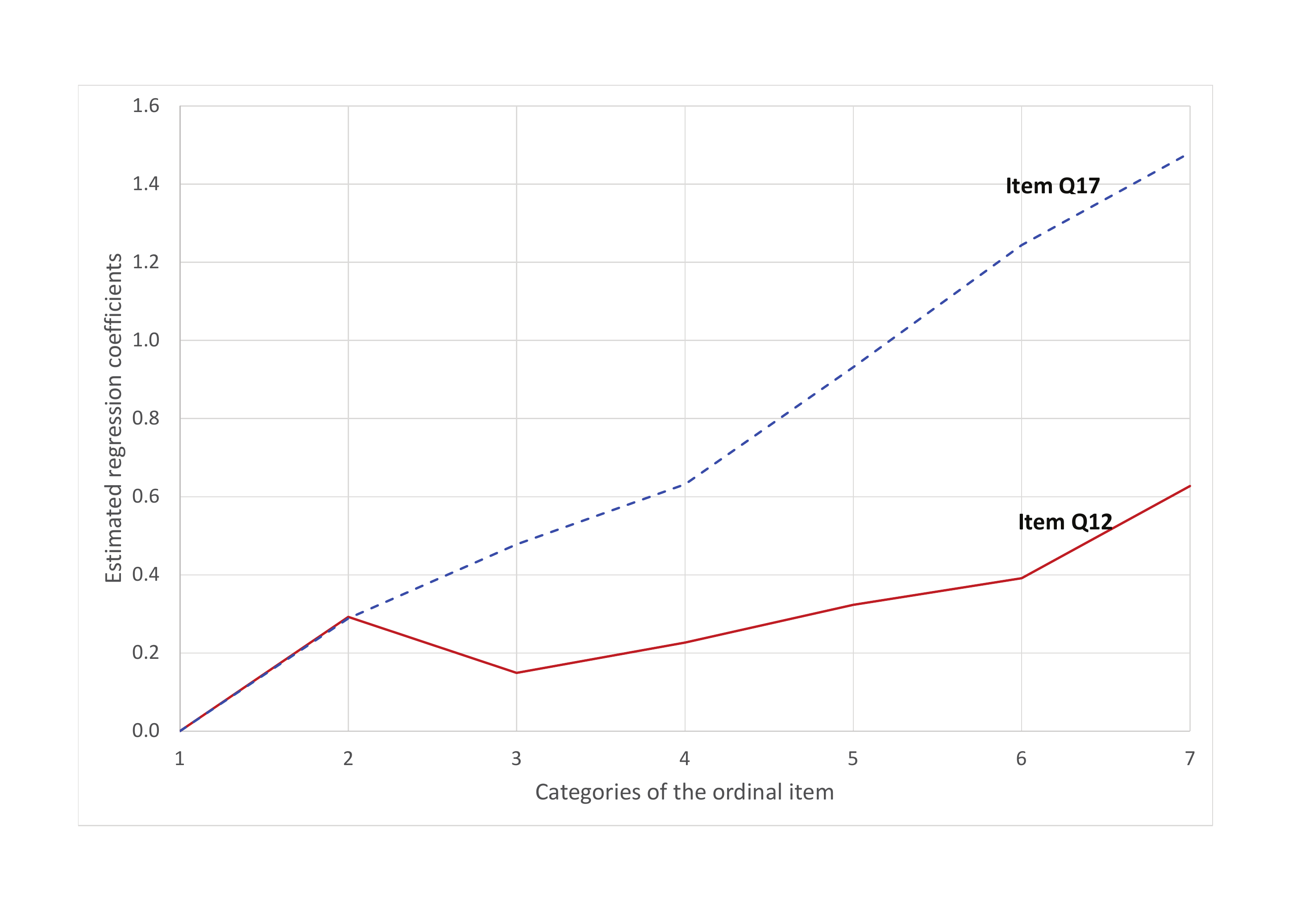}}
\vspace{-2em}
\caption{\label{fig1}Estimated regression coefficients for items $Q12$ and $Q17$ (dummy coding).}
\end{figure}

The imputation method outlined in  Section \ref{sec:imputation} preserves the 7 point scale of the ordinal items. A simple way to specify  the effect of an ordinal predictor in a regression model is to treat category codes as scores and include a single regression coefficient. This specification imposes a strong linearity assumption which may be relaxed by dummy coding, where each category is represented by an indicator variable (except for the reference category).  For example, fitting model (\ref{eq:model}) with dummy coding for ordinal items $Q12$ and $Q17$ (first category as reference), item $Q12$ does not show a linear effect (Figure \ref{fig1}).

This example makes clear that, in our case, it is advisable to begin with dummy coding for all items, and to apply data driven methods to choose a suitable specification for each item. Indeed, dummy coding yields a flexible but not parsimonious model, since it entails estimating $6 \times 20=120$  parameters for the ordinal predictors. Therefore, we need regularization methods allowing us to retain the flexibility of the dummy coding specification, while ensuring model parsimony. 

Regularization methods for ordinal predictors \cite{Gertheiss:10, Tutz:16} have a twofold aim: (\textit{i}) investigating which variables should be included in the model; (\textit{ii}) investigating which categories of an ordinal predictor should be distinguished and which can be collapsed. 
In the presence of $K$ ordinal predictors, each having $C_k$ categories, \cite{Gertheiss:10} suggest to implement the \emph{lasso} with the following $L_1$-penalty term:
\begin{equation}
\label{eq:penalty}
J(\bm \gamma) = \sum_{k=1}^{K}\sum_{c = 2}^{C_k} w_{kc} \lvert \gamma_{kc} - \gamma_{k, c-1} \rvert,
\end{equation}
where $\gamma_{kc}$ is the coefficient of the dummy variable identifying the $c$-th category of the $k$th predictor (with $\gamma_{k1}=0$ for the baseline category) and $w_{kc}$ are weights allowing for adaptive \emph{lasso}.
This approach can be applied to select all items of the PRODID questionnaire, including both ordinal and binary items, since a binary predictor is just an ordinal predictor with $C_k=2$. 

In order to exploit existing software for regularization, we use the \emph{backward difference coding}, also known as  \emph{split coding } \citep{Walter:87, Gertheiss:10}. Such a procedure allows a reparameterisation of the model in equation \eqref{eq:model} in terms of the new parameters for the ordinal predictors:
\begin{equation}\label{eq:splitpar}
\tilde{\gamma}_{kc}= \gamma_{kc} - \gamma_{k,c-1}
\end{equation}
and, in turn, the estimation of model parameters by means of a standard \emph{lasso}-type optimisation. Original dummy coefficients are simply obtained by back-transforming $\tilde{\gamma}_{kc}$; that is $\gamma_{kc} = \sum_{r = 1}^c \tilde{\gamma}_{kr}$. Note that split coding does not affect binary items, so that for such items $\tilde{\gamma}=\gamma$.

The weights $w_{kc}$ in equation \eqref{eq:penalty} are chosen as suggested by \cite{Zou:06}, yielding an adaptive \emph{lasso} procedure for parameter estimation with the following penalty term: 
\begin{equation}
\label{eq:lasso2}
J(\bm \tilde\gamma) = \sum_{k=1}^{K}\sum_{c = 2}^{C_k} \frac{1}{\lvert \hat{\tilde \gamma}_{kc} \rvert} \lvert \tilde{\gamma}_{kc}\rvert,
\end{equation}
with $\hat{\tilde \gamma}_{kc}$ denoting the Ordinary Least Squares estimate of model parameter $\tilde{\gamma}_{kc}$.

In practice, in our application, we used the command \texttt{lasso2} included in the \texttt{lassopack} module of Stata \citep{Ahrens:18} for parameter estimation.
In the following, we outline the regularization algorithm as implemented in this procedure, which relies on \cite{Belloni:12}.
In particular, the regularization procedure of \texttt{lasso2} minimizes the following penalized criterion:
\begin{equation}\label{eq:rss}
Q(\bm \tilde\gamma)=\frac{1}{n} RSS(\bm \alpha, \bm \delta, \bm \tilde\gamma) + \frac{\lambda}{n} J(\bm \tilde\gamma),
\end{equation}
where $n$ is the sample size, $\bm \alpha, \bm \delta$, and $\bm \tilde\gamma$ denote the model parameters, $RSS(\bm \alpha, \bm \delta, \bm \tilde\gamma)$ is the residual sum of squares corresponding to model \eqref{eq:model} (after split-coding the $\bm q_j$ variables), $\lambda$ is the overall penalty parameter, and $J(\bm \tilde\gamma)$ is the penalty term in equation \eqref{eq:lasso2}.  To minimize the objective function (\ref{eq:rss}), \texttt{lasso2} exploits a coordinate descent algorithm \citep{Fu:98}.

As far as the penalty parameter $\lambda$ in equation (\ref{eq:rss}) is concerned, this is chosen by minimizing the extended BIC index (\textit{EBIC}) proposed by \cite{Chen:08} and implemented in the \texttt{lasso2} procedure as follows:
\begin{equation}
\label{eq:ebic}
EBIC     = n  \log(RSS/n) + s  \log(n)+ 2s  \log(p).
\end{equation}
In the equation above, $s$ is the number of parameters in the fitted model, while $p$ is the number of parameters in the full model. Note that $EBIC$ is equal to the standard $BIC$ plus the term $2s  \log(p)$.

It is worth to notice that the \texttt{lasso2} procedure relies on a standard linear model, while the  model of interest appearing in equation \eqref{eq:model} is a linear random intercept model. We tried a specific procedure for linear mixed models, namely the \texttt{lmmlasso} package of  \texttt{R} \citep{Schelldorfer:11, Groll:14}, but we encountered computational difficulties due to the large size of the data set. However, the random effects are expected to have a little role in the regularization process for the predictors. 
Moreover, in order to reduce the bias induced by penalization, it is in general advisable to refit the model using only the selected predictors \citep{Gertheiss:10,Belloni:13}. Thus, we use the computationally efficient algorithm of \texttt{lasso2} to perform variable selection, then we fit the random intercept model (\ref{eq:model}) on the selected predictors.

\section{Combining variable selection and multiple imputation}
\label{sec:combine}
Our case study raises the additional issue of combining variable selection with multiple imputation. \cite{Zhao:17} review different approaches, highlighting that many aspects are still not fully explored. We choose the common approach of applying variable selection on each imputed data set and then combine the results. Specifically, we devise the following strategy:
\begin{enumerate}
    \item generate $M$ imputed data sets using MICE, as described in Section \ref{sec:imputation};
    \item for each imputed data set, perform variable selection  using adaptive lasso for ordinal predictors, as outlined in Section \ref{sec:lasso};
    \item retain the predictors selected in at least $k\%$ of the $M$ imputed data sets; specifically we choose a threshold of $50\%$ as in setting $S2$ of \cite{Wood:08}, which performed well in their simulation study;
    \item for each  imputed data set, fit the linear random intercept model (\ref{eq:model}) including the retained predictors;
    \item combine the $M$ vectors of estimated coefficients and the corresponding standard errors exploiting Rubin's rules \citep{Little:02};
    \item perform statistical tests on the regression parameters to refine the set of retained predictors;
    \item repeat steps (d)--(f) to choose the final model.
\end{enumerate}     
This strategy allows us to select the ordinal predictors while giving proper standard errors, namely accounting for both the hierarchical structure of the data and the uncertainty due to multiple imputation. Step (f) is advisable since it allows us to exploit the availability of proper standard errors to refine variable selection.

\section{Results}
\label{sec:results1}

The strategy outlined in  Section \ref{sec:combine} is applied to the case study on student ratings presented in Section \ref{sec:data}, which raises problems of missing data and selection of ordinal predictors.

The model of interest is the random intercept model (\ref{eq:model}). At level 1, the model includes student predictors $\bm x_{ij}$, which are centered around their cluster average in order to interpret the associated parameters as within effects \citep{Snijders:12}. At level 2, the model includes teacher and course predictors from administrative archives $\bm z_j$ (fully observed), and  teacher practices and beliefs $\bm q_j$ (subject to missing). The vector $\bm q_j$  contains dummy variables for $10$ binary items and for $20$ ordinal items. Adopting the backward-difference coding of Section \ref{sec:lasso}, the total number of parameters for the $20$ ordinal items is $6 \times 20=120$. 

The imputation step is carried out with MICE as described at the end of Section \ref{sec:imputation}. We generate $M=10$ imputed data sets.

The variable selection procedure begins by applying the regularization method described in Section \ref{sec:lasso} to each imputed data set, in order to select binary and ordinal items from the PRODID questionnaire,  while the other predictors are included in the model without penalization. We retain the predictors selected in at least $50\%$ of the imputed data sets, namely 5 binary items and 13 ordinal items. For each ordinal item $k$, the procedure selects only a subset of the $\tilde \gamma_{kc}$ parameters defined in equation (\ref{eq:splitpar}), implying collapsing of categories. Overall, the regularization procedure reduces the number of parameters $\tilde \gamma_{kc}$ from $120$ to $26$. 

The analysis proceeds by fitting  model  (\ref{eq:model}) with the retained predictors on $M=10$ imputed data sets and combining the results with the Rubin's rules. The model is fitted by maximum likelihood using the \texttt{mixed} and \texttt{mi} commands of Stata   \citep{Stata:17}. 
The variable selection procedure is refined using statistical tests based on the standard errors obtained by Rubin's rules, as suggested in step (f) of Section \ref{sec:combine}. After this step, the  final model includes the binary item $Q02$ and the ordinal items $Q12$,  $Q15$, $Q17$,  $Q27$. Table \ref{tab:results} reports final model results, while descriptive statistics of model variables are reported in Tables \ref{tab:descriptive} and \ref{tab:descriptive_MI} in the Appendix.
As shown in Table \ref{tab:results}, the selection procedure on ordinal predictors yielded predictor-specific collapsing of categories.
For example, for item $Q12$ Table \ref{tab:results} reports two coefficients corresponding to the following collapsed categories: $\{1,2,3,4\}$ (baseline), $\{5,6\}$, $\{7\}$. This means that the effect of item $Q12$ on the response variable is constant within the  collapsed categories. This result is due to the selection procedure, which retained for predictor $Q12$ two out of six parameters in equation (\ref{eq:splitpar}), specifically $\tilde \gamma_{12,5}$ and $\tilde \gamma_{12,7}$. 
Due to backward-difference coding, the parameters of the ordinal items represent contrasts between adjacent categories, thus $\hat {\tilde \gamma}_{12,5} =0.3204$ is the effect of passing from category $\{1,2,3,4\}$ to category  $\{5,6\}$, while $\hat {\tilde \gamma}_{12,7} =0.2688$ is the effect of passing from category $\{5,6\}$ to category $\{7\}$. The sum of the two parameters, $0.3204+0.2688=0.5892$, is the effect of passing from category $\{1,2,3,4\}$ to category  $\{7\}$.

\vspace{1 em}
\textbf{[Table 1 here]}
\vspace{1 em}

We briefly comment on the main finding about the effects of teacher characteristics on their ability to motivate students. Table \ref{tab:results} shows that older teachers and female teachers obtain on average lower ratings, controlling for the remaining covariates. The contribution of external experts ($Q02$) has a positive effect; this is the only item retained by the selection process out of the ten items about practices. As for beliefs, only four out of 20 items are significantly related to the ratings. In particular, ratings tend to be higher for teachers who feel that teaching is an exciting experience ($Q12$) and student opinions are a key indicator of course quality ($Q17$). On the contrary, ratings tend to be lower for teachers who think that cooperation among students helps learning ($Q15$) and teachers interested in discussing didactic methods with colleagues ($Q27$). 

In order to assess  the overall contribution of teacher practices and beliefs to explain differences in the ratings among courses, we compare the residual level 2 variance under different model specifications. In particular, fitting model (\ref{eq:model}) without any predictor yields an estimated level 2 variance  $\hat \sigma^2_u= 1.3320$, which reduces to  1.2306 ($-8$\%) after introducing all the predictors except teacher practices and beliefs.  The final model gives $\hat \sigma^2_u=1.0012$, corresponding to a further reduction of residual level 2 variance of about $19$\%. Thus, teacher practices an beliefs are the most relevant observed factors in explaining differences in the ratings among courses.

To evaluate the performance of the imputation procedure, the last two columns of Table \ref{tab:results} report the  diagnostic measures  $FMI$ and $RE$, which are derived from  the decomposition of the total sampling variance $V_T$ of an estimator \citep{Enders:10}:
\begin{equation}\label{eq:MIVar}
 V_T= V_W+ V_B+V_B/M
\end{equation}
where $ V_B= \frac{1}{M-1}\sum _{m=1}^M \left(\hat \beta_m-\overline{\hat \beta}\right)^2$ is the  between-imputation variance,  while   $V_W=\frac{1}{m}\sum _{m=1}^M SE(\hat \beta_m)^2$ is the within-imputation variance, with $SE(\hat \beta_m)$ denoting the standard error  obtained from the $m$-th  imputed data set.
The index $FMI$ (Fraction of Missing Information) is used to quantify the influence of multiple imputation on the sampling variance of a parameter estimate:
\begin{equation}\label{eq:FMI}
    FMI=\frac{ V_B+ V_B/M}{V_T}
\end{equation}
On the other hand, the index RE  (Relative Efficiency) is the relative efficiency for using a finite number of imputations ($M=10$ in our case) versus the theoretically optimal infinite number of imputations:
\begin{equation}\label{eq:RE}
    RE=\left(1+\frac{FMI}{M}\right)^{-1}
\end{equation}
As regards for level 1 predictors, Table \ref{tab:results} shows values of $FMI$ near zero and values of $RE$ near one. Indeed, level 1 predictors are fully observed and cluster-mean centered, so they are not affected by imputations of level 2 predictors. Fully observed level 2 predictors (i.e. teacher and course characteristics) are little affected by imputations, showing $FMI$ between $0.01$ and $0.17$, and relative efficiency close to 1.  For imputed level 2 predictors (i.e. teacher practices and beliefs) $FMI$ ranges from $0.27$ to $0.49$, with a mean value of $0.40$, indicating that on average $40\%$ of the sampling variance is attributable to missing data, which is lower than the fraction of missing values in the data set (about $50\%$). This points out a favourable  trade-off between the increase of sampling error due to imputations and its reduction due to data augmentation. Moreover, the relative efficiency for imputed predictors ranges from $0.953$ to $0.973$, suggesting that $M=10$ imputations are acceptable to obtain a satisfactory level of efficiency.  

\section{Concluding remarks}
\label{sec:conclusion}

In this paper we considered a complex analysis involving  a multilevel model with many level 2 ordinal and binary predictors affected by a  high rate of missing values. We proposed a strategy to jointly handling missing values and selecting categorical predictors. The proposed solution combines existing methods in an original way to solve the specific problem at hand, but it is generally applicable to settings requiring to select categorical predictors affected by missing values. 
Specifically, we handled missing data using Multiple Imputation by Chained Equations. This allowed us to retain all the observations and analyze the data under the MAR assumption instead of the unrealistic MCAR assumption. The MAR assumption seems plausible since the imputation model exploits all the information from level 1 and level 2 observed values. The ordinal and binary predictors were selected using an ad hoc regularization method, namely the \emph{lasso} for ordinal predictors. The regularization procedure induces a data-driven specification of the relationship between the response and the ordinal predictors, by collapsing the categories. This method can be easily extend to handle also nominal predictors \citep{Tutz:16}. Regularization was applied separately on each imputed data set and the results were combined retaining the parameters selected in at least half of the imputed data sets.  Finally, the random effect model of interest was fitted including the chosen predictors. The uncertainty due to imputation is accounted by Rubin's rules. The proposed procedure allowed us to specify the model in a flexible, though parsimonious way.

The results obtained with the final model pointed out that some teacher practices and beliefs are significantly related to ratings about teacher ability to motivate students. 

The complexity of the case study, especially in terms of number of observations and  number of categorical variables affected by missing values, forced us to rely on computationally low demanding algorithms. The solution of computational issues would allow us to extend the proposed approach in several ways. For example, in the imputation step MICE could be replaced by Joint Modelling  \citep{Goldstein:14, Quartagno:16} or by the latent class approach \citep{Vidotto:18}, which give valid inferences for a wider set of specifications of the model of interest, including non linear effects and interactions. 

Combining model selection with multiple imputation is an open issue \citep{Zhao:17}. We devised a simple strategy to face a computationally demanding setting, but it would be interesting to explore other approaches. 

\section*{Acknowledgements}
The authors gratefully acknowledge the support of the University of Padua project Advances in Multilevel and Longitudinal Modelling, principal investigator Omar Paccagnella, grant no. SID2016

\newpage
\section*{Appendix}

\textbf{[Tables 2 and 3 here]}

\begin{table}
\caption{\label{tab:results}Multiple imputation estimates: random intercept model for student satisfaction on teacher  ability to motivate students.}
\centering
\fbox{%
\begin{tabular}{*{6}{lrrrrr}}
\hline
\emph{Covariates} &\emph{Coeff}& \emph{SE}& \emph{p-value}& \emph{FMI}$\dag$&\emph{RE}$\dag$\\ \hline
Constant   &   7.7446 &  0.2628  &    0.000   &  0.1817  &     0.9822\\   
\multicolumn{1}{l}{\emph{Student characteristics (lev 1)}} &&&&&\\ 
Female               &  -0.0515  & 0.0188  &    0.006   &  0.0000  &     1.0000\\
Age                  &   0.0479  & 0.0029  &    0.000   &   0.0000  &    1.0000\\
High School grade    &   0.0072  & 0.0008  &    0.000   &  0.0000  &     1.0000\\
Enrollment year      &  -0.0918  & 0.0295  &    0.002   &  0.0002  &    1.0000\\
Regular enrollment   &  -0.1697  & 0.0485  &    0.000   &  0.0000  &     1.0000\\
Passed exams         &   0.1874  & 0.0385  &    0.000   &  0.0001  &     1.0000\\
\multicolumn{1}{l}{\emph{Course characteristics  (lev 2)}} &&&&&\\ 
Compulsory course    &   -0.2169&    0.0431 &      0.000   &  .01332  &  0.9987\\
\multicolumn{2}{l}{School }&&&&                 \\
\hspace{3ex}Agronomy and Veterinary& - &- &-&-&-\\
\hspace{3ex}Social Sciences &   0.0516 &   0.1505  &     0.732   &  0.1084  &     0.9893\\
\hspace{3ex}Engineering     &  -0.3209 &   0.1279  &     0.012   &  0.0785  &     0.9922\\
\hspace{3ex}Psychology      &   0.2142 &   0.1619  &     0.186   &  0.0526  &     0.9948\\
\hspace{3ex}Sciences        &   0.0444 &   0.1340  &     0.741   &  0.1662  &     0.9837\\
\hspace{3ex}Humanities      &   0.2290 &   0.1393  &     0.101   &  0.1544  &     0.9848\\
\multicolumn{1}{l}{\emph{Teacher characteristics (lev 2)}} &&&&&\\                                                
 Female              &   -0.1377 &  0.0793   &   0.083       & 0.0987  &     0.9902\\
 Age (years)                 &   -0.0157 &  0.0039   &   0.000     & 0.1266  &     0.9875\\
\multicolumn{1}{l}{\emph{Teacher practices (lev 2)}}&&&&&\\ 
         Q02 External contributors  & 0.2645   & 0.0991    &  0.010 &  0.4287  &     0.9589\\
\multicolumn{1}{l}{\emph{Teacher beliefs (lev 2)}} &&&&&\\ 
\multicolumn{1}{l}{Q12 Teaching exciting experience} &&&&&\\
\hspace{3ex}  \{1,2,3,4\}& - &- &-&-&-\\
 \hspace{3ex} \{5,6\} &  0.3204 &   0.1236   &    0.012   &  0.4194  &     0.9598\\
 \hspace{3ex} \{7\} &  0.2689 &   0.0948   &    0.006   &  0.3140  &     0.9696\\
\multicolumn{1}{l}{Q15 Student cooperation useful} &&&&&\\
\hspace{3ex}  \{1,2,3,4,5\}& - &- &-&-&-\\
 \hspace{3ex} \{6,7\}& -0.2338 &   0.0931   &    0.015   &  0.4455  &     0.9574\\
\multicolumn{1}{l}{Q17 Student opinions relevant} &&&&&\\
\hspace{3ex}  \{1,2,3,4\}& - &- &-&-&-\\
 \hspace{3ex} \{5\} &  0.3891 &   0.1227   &    0.002   &  0.4209  &     0.9596\\
 \hspace{3ex} \{6\} &  0.3190 &   0.1308   &    0.019   &  0.4890  &     0.9534\\
\hspace{3ex}  \{7\} &  0.2340 &   0.1182   &    0.050   &  0.2705  &     0.9737\\
\multicolumn{1}{l}{Q27 Discuss teaching methods} &&&&&\\
\hspace{3ex}  \{1,2\}& - &- &-&-&-\\
\hspace{3ex} \{3,4,5,6,7\} & -0.2974 &   0.1055   &    0.006   &  0.3781  &     0.9636\\
\multicolumn{1}{l}{\emph{Residual variances}}&&&&&\\ 
  $\sigma^2_e$ (level 1)  & 3.3971&    &  &  &   \\
   $\sigma^2_u$ (level 2)  & 1.0012 &    &  &  &    \\ \hline
 \multicolumn{6}{l}{$\dag$ \emph{FMI} defined in (\ref{eq:FMI}), \emph{RE} defined in (\ref{eq:RE})}\\ 
\end{tabular}}
\end{table}

\begin{table}
\caption{\label{tab:descriptive}Descriptive statistics of fully observed variables (56775 ratings, 1016 courses)}
\centering
\fbox{%
\begin{tabular}{*{6}{lrrrr}}
\hline
\emph{Variables} &\emph{Mean}& \emph{sd}& \emph{min}& \emph{max}\\ \hline
\multicolumn{1}{l}{\emph{Outcome (lev 1)}} &&&&\\ 
Student rating on teacher ability   &  7.387  & 2.163   & 1      & 10      \\
\multicolumn{1}{l}{\emph{Student characteristics (lev 1)}} &&&&\\ 
Female               &     0.511  &      0.500 &      0&       1\\ 
Age                  &    20.511  &      2.962 &     17&      78\\ 
High School grade    &    80.729  &     11.817 &     60&     100\\ 
Enrollment year      &    1.688   &     0.786  &      1&       3\\
Regular enrollment   &     0.967  &      0.178 &      0&       1\\
Passed exams         &    0.608   &     0.259  &      0&       1\\

\multicolumn{1}{l}{\emph{Teacher characteristics (lev 2)}} &&&&\\                                                
 Female           &   0.324   &    0.468    &  0   &   1\\
 Age (years)      &   50.638  &    9.442    &  32   &   70  \\

\multicolumn{1}{l}{\emph{Course characteristics  (lev 2)}} &&&&\\ 
Number of ratings (per course) &79.563& 58.902  &5        &442 \\
Compulsory course              &0.296 &  0.457  & 0       & 1   \\ 
School&&&& \\
\hspace{4.5ex}Agronomy and Veterinary &0.109 &    &        &    \\
\hspace{4.5ex}Social Sciences         &0.113 &    &        &  \\ 
\hspace{4.5ex}Engineering             &0.237 &    &        &  \\ 
\hspace{4.5ex}Psychology              &0.075 &    &        &  \\ 
\hspace{4.5ex}Sciences                &0.256 &    &        &  \\ 
\hspace{4.5ex}Humanities              &0.210 &    &        &  \\ 
\end{tabular}}
\end{table}

\begin{table}
\caption{\label{tab:descriptive_MI}Statistics of Teacher practices and beliefs (1016 teachers).}
\centering
%\begin{threeparttable}
\fbox{%
\begin{tabular}{*{5}{lrrrr}}
\hline
\emph{Item} &\emph{\% missing} &\emph{Category}&\multicolumn{2}{c}{\emph{Relative Frequency}}\\ \cline{4-5}
                            &&& \emph{Observed}& \emph{MI average} $\dag$  \\ \hline
Q02 External contributors & 46.75            &  &0.340   &0.344   \\    
Q12 Teaching exciting experience  &47.15&&&\\     
                                        &&1  & 0.011 &0.040 \\   
                                        &&2  &0.017  &0.028 \\   
                                        &&3  &0.039  &0.046 \\   
                                        &&4  & 0.069 &0.067 \\   
                                        &&5  &0.229  &0.212 \\   
                                        &&6  &0.330  &0.307 \\   
                                        &&7  &0.305  &0.300 \\   
Q15 Student cooperation useful &48.03&&&\\
                                        &&1  & 0.019 &0.034\\   
                                        &&2  & 0.040 &0.052\\   
                                        &&3  & 0.076 &0.093\\   
                                        &&4  & 0.142 &0.145\\   
                                        &&5  & 0.237 &0.214\\   
                                        &&6  & 0.307 &0.288 \\  
                                        &&7  & 0.180 &0.174\\   
 Q17 Student opinions relevant &47.24&&&\\
                                       &&1   & 0.017 &0.024  \\  
                                       &&2   & 0.045 &0.059 \\   
                                       &&3   & 0.065 &0.069 \\  
                                       &&4   & 0.151 &0.159 \\
                                       &&5   & 0.237 &0.224\\   
                                       &&6   & 0.289 &0.277\\    
                                       &&7   & 0.196 &0.189 \\   
Q27 Discuss teaching methods &47.83&&&\\
                                       &&1   & 0.093  &0.127\\   
                                       &&2   & 0.076  &0.087 \\  
                                       &&3   & 0.113 &0.116 \\   
                                       &&4   & 0.155 &0.139 \\     
                                       &&5   & 0.155 &0.142\\    
                                       &&6   & 0.226 &0.211 \\     
                                       &&7   & 0.183 &0.178 \\   
                                      
 \end{tabular}}
 %\begin{tablenotes}
  % \item[*] 
   \noindent $\dag$ MI averages are obtained on 10 data sets imputed by MICE (see Section \ref{sec:imputation}).
  %\end{tablenotes}
  %\end{threeparttable}
\end{table}

\end{document}